\documentclass[aps,prb,twocolumn,groupedaddress,showpacs]{revtex4}

\usepackage{graphicx}
\usepackage{dcolumn}
\usepackage{bm}
\usepackage{amsmath,amssymb}
\usepackage{color}

\def\Vec#1{\mbox{\boldmath $#1$}}
\def\rc{{\rm c}}

\begin{document}


\title{Field-induced spin nematic Tomonaga-Luttinger liquid of the $S=1/2$ spin ladder system with the anisotropic ferromagnetic rung interaction}



\author{T\^oru Sakai$^{1,2}$, Ryosuke Nakanishi$^1$, Takaharu Yamada$^1$, Rito Furuchi$^1$, 
Hiroki~Nakano$^1$, Hirono Kaneyasu$^1$, Kiyomi Okamoto$^{1}$ and Takashi Tonegawa$^{1,3,4}$}

\affiliation{
$^1${Graduate School of Science, University of Hyogo, Kouto 3-2-1, Kamigori, Ako-gun, Hyogo 678-1297 Japan}\\
$^2${National Institute for Quantum Science and Technology (QST) SPring-8, Kouto 1-1-1, Sayo, Sayo-gun, Hyogo 679-5148, Japan}\\
$^3${Professor Emeritus, Kobe University, Kobe 657-8501, Japan} \\
$^4${Department of Physics, Graduate School of Science, Osaka Metropolitan University, Sakai 599-8531, Japan}\\
}


\date{\today}

\begin{abstract}

The $S=1/2$ quantum spin ladder system with the anisotropic ferromagnetic 
exchange interaction on the rung under magnetic field is investigated using 
the numerical diagonalization and the density 
matrix renormalization group (DMRG) analyses. 
It is found that the nematic-spin-correlation-dominant Tomonaga-Luttinger 
liquid (TLL) appears in some high magnetic field. 
It is included in the TLL phase where the two-magnon 
bound state is realized. 
For some suitable parameters, after the field-induced phase transition 
from this two-magnon-bound TLL phase to the single-magnon 
TLL one, the re-entrant transition to the two-magnon-bound TLL phase 
occurs, which is confirmed by the magnetization curves by the DMRG. 
Several phase diagrams on the plane of the coupling anisotropy 
versus the magnetization and the magnetic field are presented. 
The present result is a proposal of the candidate system which 
exhibits the spin nematic phase without the biquadratic interaction 
or the frustration. 

\end{abstract}


\maketitle


\section{Introduction}

The spin nematic state\cite{nematic1,nematic2} has attracted a lot of interest in the research field of the 
quantum spin systems and the strongly correlated electron systems. 
It is the long-range quadrapole order of spins by forming the two-magnon bound state. 
In the one-dimensional case, due to the strong quantum fluctuations,
the nematic long-range order is reduced to the nematic quasi long-range order,
which should be called spin nematic Tomonaga-Luttinger liquid (TLL).
Namely, the quadrapole correlation function decays in the power-law
in the spin nematic TLL phase.

It was shown that, in the $S=1/2$
ladder with different leg interactions and some anisotropies,
two kinds of spin nematic TLL phases appeared, by using numerical
diagonalization calculations, the density matrix renormalization
group (DMRG) method, and perturbation calculations.\cite{tonegawa}
The spin nematic TLL phase was found in the simple $S=1$ chain with the $XXZ$
and on-site anisotropies.\cite{solyom-ziman,schulz,nijs,chen-hida,sakai2022}
The $S=1$ bilinear and biquadratic chain was also theoretically predicted
to exhibit the spin nematic TLL phase by several methods; 
the perturbation,\cite{chubukov} the bosonization,\cite{solyom}
the numerical exact diagonalization,\cite{lauchli,sakai2022} 
the field theory,\cite{senthil} the DMRG\cite{manmana}
 and the infinite matrix product state analysis.\cite{mao}  
Furthermore, the spin nematic TLL phase was revealed to occur
 in the $S=3/2$ bilinear and biquadratic model.\cite{chubukov, fridman} 
The spin frustration is another important mechanism to induce
 the spin nematic phase.\cite{chandra}  
In order to explain the spin liquid like behavior
of the $S=1$ triangular magnet NiGa$_2$S$_4$,\cite{nakatsuji} 
the spin nematic phase was proposed.\cite{tsunetsugu, penc, shenoy, park} 
The frustrated spin chain which has the ferromagnetic
nearest- and the antiferromagnetic next-nearest-neighbor exchange interactions, 
are one of popular models to exhibit the spin nematic TLL phase. 
The external magnetic field induced spin nematic TLL phase
was predicted to occur in the $S=1/2$ chain with 
the ferromagnetic nearest- and the antiferromagnetic next-nearest-neighbor exchange interactions 
by the bosonization,\cite{chubukov2}
the numerical exact diagonalization,\cite{vekua,sudan} the DMRG\cite{sudan,hikihara,sato1} 
and the field theory.\cite{starykh,sato1}

Several experimental methods to detect the spin nematic behavior were 
theoretically proposed, for example the NMR,\cite{sato2,sato3,shindou,smerald1,smerald2}
the inelastic neutron scattering,\cite{smerald3} 
the $\mu$SR,\cite{smerald1} and the ESR.\cite{furuya} 
One of the suitable candidate materials to exhibit the spin nematic behavior 
is LiCuVO$_4$ which is the $S=1/2$ quasi-one-dimensional quantum spin system 
with the ferromagnetic nearest- and the antiferromagnetic 
next-nearest-neighbor exchange interactions.\cite{buttgen} 
The NMR measurements\cite{buttgen,orlova} 
on this compound under high magnetic field detected an 
evidence of the possible spin nematic order, as well as the magnetocaloric 
effect measurement.\cite{gen} 
The NMR experiment\cite{mendels} on the similar compound 
LiCuSbO$_4$\cite{dutton} also observed the spin nematic order like behavior. 
Since the iron-based superconductors\cite{hosono,stewart,si,dai,dagotto} 
were discovered, the spin nematic physics on the two dimensional systems 
\cite{lauchli2,nic1,capponi,shindou2,harriger,hu, tanaka1}
 have been studied extensively, 
including the bilayer systems.\cite{yokoyama,hikihara2,tanaka2}

In most theories of the spin nematic behavior which have been proposed so far, 
the mechanism is based on the biquadratic interaction or the spin frustration. 
In this paper we propose a simple theoretical model that exhibits 
the field-induced spin nematic TLL, without either the biquadratic interaction 
or the frustration. 
It is the $S=1/2$ spin ladder system with the anisotropic ferromagnetic 
rung exchange interaction under magnetic field. 
In the previous work\cite{sakai1} the numerical diagonalization and 
the DMRG calculation indicated that the present model with the same amplitude 
between the antiferromagnetic leg and the ferromagnetic rung interactions 
gives rise to the field induced spin nematic TLL phase.
In the present work the critical exponent analysis indicates that 
the spin nematic correlation dominant region and the SDW correlation dominant region
appear in the two-magnon-bound TLL phase. 
In addition we present several phase diagrams not only in 
the anisotropy-magnetization plane, but also in 
the anisotropy-external field plane, 
even for different amplitudes between the leg and rung interactions. 
The magnetization curves calculated by the DMRG are also presented for 
several typical cases.

\section{Model}
We consider the magnetization process 
of the $S=1/2$ Heisenberg spin ladder with the anisotropic ferromagnetic 
rung exchange interaction.  
The Hamiltonian is given by 
\begin{eqnarray}
\label{ham}
&{\cal H}&={\cal H}_0+{\cal H}_Z, \\
&{\cal H}_0& = J_1 \sum _{\alpha=1}^2 \sum_{j=1}^L \Vec{S}_{\alpha,j}\cdot \Vec{S}_{\alpha,j+1} \nonumber \\
&& +J_{\rm r} \sum_{j=1}^L\left[S_{1,j}^xS_{2,j}^x + S_{1,j}^yS_{2,j}^y 
 + \lambda S_{1,j}^zS_{2,j}^z  \right] \\
&{\cal H}_Z& =-H\sum _{\alpha=1}^2 \sum_{j=1}^L S_{\alpha,j}^z,
\end{eqnarray}
where $\lambda$ is an anisotropy parameter of the ferromagnetic rung exchange interaction 
and $H$ is the external magnetic field. 
The ferromagnetic rung interaction constant $J_{\rm r}$ is set to be $-1$.
We consider the case of the antiferromagnetic leg interaction $J_1>0$ and 
the Ising-like anisotropy $\lambda >1$ of the ferromagnetic rung interaction. 
For the length $L$ system, 
the lowest energy of ${\cal H}_0$ in the subspace where 
$\sum_i\sum _j S_{i,j}^z=M$ is denoted by $E(L,M)$. 
The reduced magnetization $m$ is defined by $m=M/M_{\rm s}$, 
where $M_{\rm s}$ denotes the saturation of the magnetization, 
namely $M_{\rm s}=L$. 
The energies $E(L,M)$ are calculated by the Lanczos algorithm under the 
periodic boundary condition ($ \Vec{S}_{i,L+1}=\Vec{S}_{i,1}$).

\section{Haldane-Neel phase boundary}

\begin{figure}
\includegraphics[width=0.85\linewidth,angle=0]{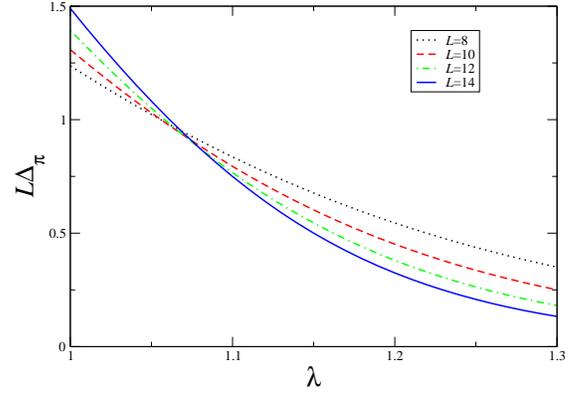}%
\caption{\label{prgm0} 
Scaled gap $L\Delta_{\pi}(L,\lambda)$ plotted versus $\lambda$ 
for $L=$8, 10, 12 and 14 in the case of $J_1=0.5$. 
}
\end{figure}

\begin{figure}
\includegraphics[width=0.85\linewidth,angle=0]{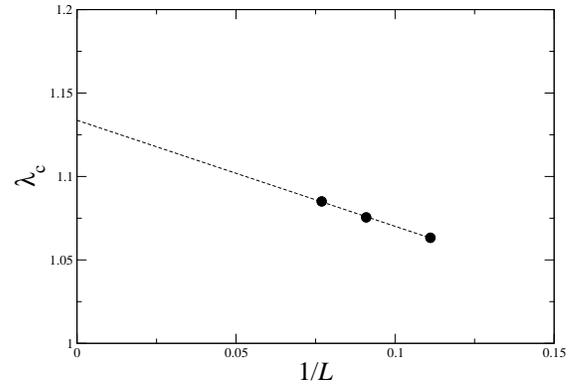}%
\caption{\label{extrap-m0} 
Extrapolation of the N\'eel-Haldane boundary at $m=0$ for $J_1 = 0.5$.
We can see that $\lambda_{{\rm c},N+1}$ converges as $1/L$.
The extrapolated value is $\lambda_{\rm c} = 1.134 \pm 0.002$.
}
\end{figure}

In the absence of the external field ($H=0$), the ground state of the 
system is in the Haldane phase with the Haldane gap for $\lambda \sim 1$, 
while in the N\'eel ordered phase for $\lambda \gg 1$. 
The phase boundary $\lambda _c$ can be estimated using the 
phenomenological renormalization group method\cite{nightingale}. 
The size-dependent critical point $\lambda_{\rc,L+1}$ is 
determined from the equation of the scaled gaps
\begin{eqnarray}
L\Delta_{\pi}(L,\lambda_\rc)=(L+2)\Delta_{\pi}(L+2,\lambda_\rc).
\label{sg}
\end{eqnarray}
where $\Delta_{\pi}(L,\lambda)$ is the lowest excitation gap with the 
wave number $k=\pi$. 
The scaled gap $L\Delta_{\pi}(L,\lambda)$ is plotted versus $\lambda$ 
for $L=$8, 10, 12 and 14  in the case of $J_1=0.5$ shown in Fig.~\ref{prgm0}. 
Assuming the size correction proportional to $1/L$, 
we estimate the phase boundary $\lambda_\rc$ in the 
infinite length limit as shown in FIG. \ref{extrap-m0}. 

\section{Two Tomonaga-Luttinger Liquid Phases}

\begin{figure}
\includegraphics[width=0.85\linewidth,angle=0]{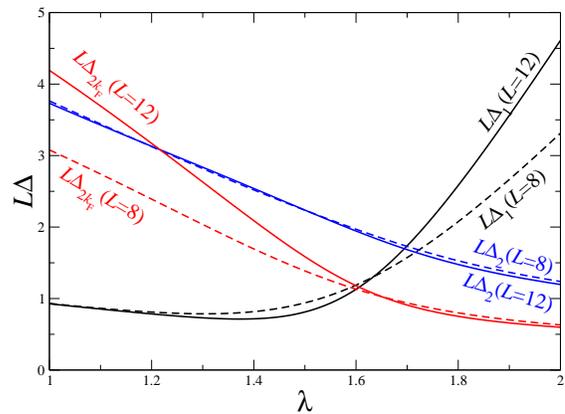}%
\caption{\label{sgmh} 
Scaled gaps $L\Delta_1$ (black curves), $L\Delta_2$ (blue curves) and $\Delta_{2k_{\rm F}}$
(red curves) plotted versus $\lambda$ for $L=8$ (dashed curves) and 12 (solid curves)
in case of $J_1=0.5$ and $m=1/2$. 
}
\end{figure}

If the magnetization process begins from the Haldane phase 
for $\lambda < \lambda_\rc$, a quantum phase transition would 
occur at the critical field $H_{\rm c1}$ where the spin gap 
vanishes, and the gapless TLL phase 
would appear for $H > H_{\rm c1}$, like the $S=1$ antiferromagnetic 
chain.\cite{affleck1,affleck2,tsvelik,sakai2,sakai3}
This TLL phase is called the conventional TLL (CTLL) phase 
where each step of the magnetization curve for the finite-size systems 
should be $\delta M=\delta S^z_{\rm tot}=1$. 
On the other hand, 
starting from the N\'eel ordered phase for $\lambda > \lambda_c$, 
the large Ising-like anisotropy
stabilizes the states 
$|\uparrow \uparrow \rangle$ and $|\downarrow \downarrow \rangle$ at each rung pair, 
which is nothing but the local two-magnon bound state.
These two states can be expressed by the $T=1/2$ pseudo-spin
with  $|T^z=+1 \rangle = |\uparrow \uparrow \rangle$
and $|T^z=-1 \rangle = |\downarrow \downarrow \rangle$.
Then the magnetization process will be similar to the
Ising-like $T=1/2$ $XXZ$ single antiferromagnetic chain.
This TLL phase is called the two-magnon-bound TLL phase, 
where each step of the magnetization curve should be $\delta M=2$. 
The gapless quasiparticle excitation is different between these two TLL phases. 
Namely, it is the single magnon excitation for the CTLL, while 
the two-magnon excitation for the two-magnon-bound TLL. 
We note that the single magnon excitation is gapped in the
two-magnon-bound TLL.
Thus the cross points between these two excitation gaps 
given by the forms
\begin{eqnarray}
\Delta _1 &=&E(L, M+1)+E(M-1)-2E(M), \\
\Delta _2 &=&E(L, M+2)+E(L,M-2)-2E(M),
\end{eqnarray}
will be the phase boundary $\lambda_\rc$ in the infinite $L$ limit 
for each magnetization $M$. 
At the half of the saturation magnetization $m=1/2$ for $J_1=0.5$, 
the scaled gaps $L\Delta _1$(black curves) and $L\Delta _2$ (blue curves) are plotted 
versus $\lambda$ for $L=8$ (dashed curves) and 12 (solid curves), respectively in Fig.~\ref{sgmh}. 
It indicates that 
$\Delta_1$ is gapless for smaller $\lambda$ but gapped for larger $\lambda$, 
while $\Delta_2$ is always gapless. 

Considering the gapless $2k_{\rm F}$ excitation, 
we can use another method to estimate the phase boundary. 
Since the wave number $2k_{\rm F}$ is different between two TLL phases, 
this excitation in the two-magnon-bound TLL phase ($2k_{\rm F}=2m\pi$) is 
gapped in the CTLL phase, while gapless in the two-magnon-bound TLL phase. 
Thus the behaviors of the $2k_{\rm F}$ excitation gap for the two-magnon-bound TLL phase $\Delta_{2k_{\rm F}}$ 
and $\Delta_1$ are just switched at the critical point. 
The scaled gap $L\Delta_{2k_{\rm F}}$ (red curve) is also plotted versus $\lambda$ for $L=$8 and 12 
in Fig.~\ref{sgmh}. 
The cross point of $\Delta_1$ and $\Delta_{2k_{\rm F}}$ would be also the phase boundary $\lambda_\rc$ 
in the infinite $L$ limit.

\begin{figure}
   \includegraphics[width=0.85\linewidth,angle=0]{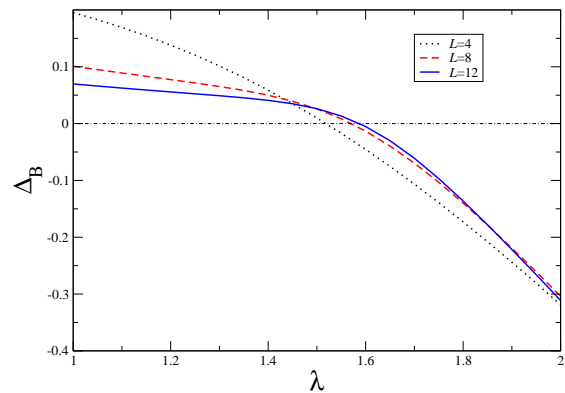}%
   \caption{\label{bindmhj105} 
   Two-magnon binding energies $\Delta_{\rm B}$ are plotted versus $\lambda$ for $J_1=0.5$ and $m=1/2$.
}
\end{figure}

\begin{figure}
   \includegraphics[width=0.85\linewidth,angle=0]{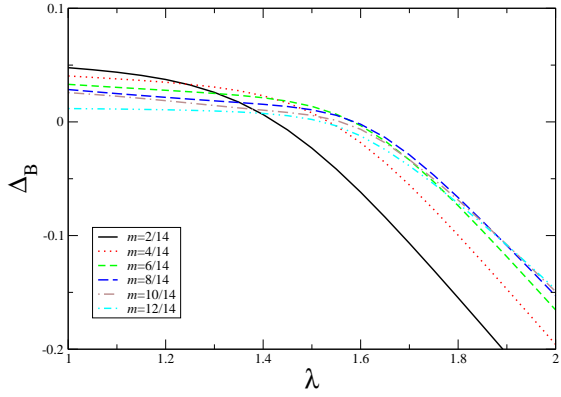}%
   \caption{\label{bindj05} 
   Two-magnon binding energies $\Delta_{\rm B}$ as functions of $\lambda$
   for various $M$ in case of $J_1 = 0.5$ and $L=14$.
}
\end{figure}

We also calculate the two-magnon binding energy\cite{parvej} defined by
\begin{eqnarray}
    \Delta_{\rm B}
    &=& \{E(L,M+2) - E(L,M)\}  \nonumber \\
    &&- 2\{ E(L,M+1) - E(L,M)\} \nonumber \\
    &=& E(L,M) + E(L,M+2) - 2E(L,M+1).
    \label{binding}
\end{eqnarray}
which should be positive in the CTLL phase while negative in the two-magnon-bound TLL phase.
Therefore the point $\Delta_{\rm B} =0$ would be the phase transition point in the $L \to \infty$ limit.
The behavior of $\Delta_{\rm B}$ for $J_ 1 =0.5$ and $m=1/2$ is shown in Fig.~\ref{bindmhj105}.
We also show the binding energies as functions of $\lambda$ for various $M$ in case of $J_0 = 0.5$ and $L=14$
in FIG. \ref{bindj05}.

\begin{figure}
\vskip1cm
\includegraphics[width=0.85\linewidth,angle=0]{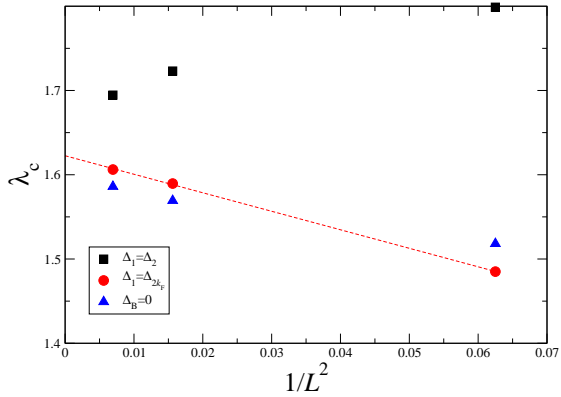}%
\caption{\label{sizedep} 
The cross points between $\Delta_1$ and $\Delta_2$ (black squares), and 
those between $\Delta_1$ and $\Delta_{2k_{\rm F}}$ (red circles) are plotted versus $1/L^2$
in case of $J_1=0.5$ and $m=1/2$. 
The points where $\Delta_{\rm B} = 0$ are also plotted (blue triangles).
The first and third points converge as $1/L$,
whereas the second poins as $1/L^2$.
The $L \to \infty$ extrapolated value of the second points is
$\lambda_{\rm c} = 1.623 \pm 0.002$.
}
\end{figure}

The behaviors of the gaps and the two-magnon binding energy
are most basic properties to distinguish the CTLL and two-magnon-bound TLL phases.
Thus the phase transition between these two phases is directly confirmed by
Figs.~\ref{sgmh} and \ref{bindmhj105}.
The cross points of $\Delta_1$ and $\Delta_2$ (black squares), 
and that of $\Delta_1$ and $\Delta_{2k_{\rm F}}$ (red circles) 
as well as the points $\Delta_{\rm B} = 0$ (blue triangles)
for $L=4, 8$ and 12 are plotted versus $1/L^2$ in Fig.~\ref{sizedep}. 
The first and third points converges with respect to $L$ almost as $1/L$,
while the second ones almost as $1/L^2$. 
Since the phase boundary depends on the magnetization $M$, 
we use the cross point of $\Delta_1$ and $\Delta_{2k_{\rm F}}$ for largest $L$ 
as the phase boundary at each $M$.

\section{Spin Correlations}

\begin{figure}
\includegraphics[width=0.85\linewidth,angle=0]{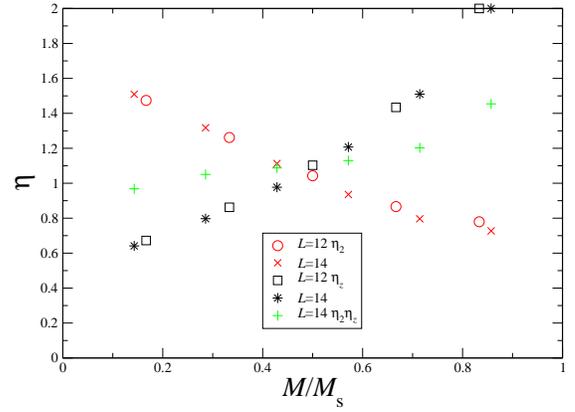}%
\caption{\label{eta} 
Critical exponents $\eta_z$  and $\eta_2$
for $L=$12 and 14 are plotted versus $M/M_s$ 
for $J_1=0.5$ and $\lambda =2.5$. 
The product $\eta_2 \eta_z$ is also plotted for $L=14$. 
}
\end{figure}

In the two-magnon-bound TLL phase the following spin correlation functions 
exhibit the power-law decay: 
\begin {eqnarray}
   &&C_z(r)
   \equiv \langle S_{\alpha,0}^z S_{\alpha,r}^z \rangle - \langle S^z \rangle^2
   \sim  \cos(2k_{\rm F}r)r^{-\eta_z}, 
   \label{sdw}\\
   &&C_2(r)
   \equiv \langle S_{1,0}^+S_{2,0}^+ S_{1,r}^-S_{2,r}^- \rangle  \sim  r^{-\eta_2}.
   \label{nematic}
\end{eqnarray}
The former corresponds to the SDW spin correlation parallel to the 
external field and the latter corresponds to the nematic 
(quadrapole) spin correlation perpendicular to the external field. 
Comparing the exponents $\eta_z$ and $\eta_2$, 
the smaller determines the dominant spin correlation. 
The area for $\eta_2 < \eta_z$ in the parameter space should be 
called the nematic correlation dominant TLL (NTLL) region, 
and that  for $\eta_2 > \eta_z$  the SDW dominant TLL (SDW$_2$TLL) region. 
According to the conformal field theory, 
these exponents can be estimated by the forms
\begin{eqnarray}
\eta_2&=&{{E(L,M+2)+E(L,M-2)-2E(L,M)}\over{E_{k_1}(L,M)-E(L,M)}}, \\
\eta_z&=&2{{E_{2k_F}(L,M)-E(L,M)}\over{E_{k_1}(L,M)-E(L,M)}},
\label{exponent}
\end{eqnarray}
for each magnetization $M$, where $k_1$ is defined as $k_1=L/2\pi$. 
The exponents $\eta_2$ and $\eta_z$ estimated for $L$=12 and 14 are plotted versus 
$M/M_s$ for $J_1=0.5$ and $\lambda$ =2.5 in Fig.~\ref{eta}. 
It suggests that the SDW correlation is dominant for small $M$, 
while the nematic one for large $M$. 
Since the cross point of $\eta_2$ and $\eta_z$ is not so strongly 
dependent on $L$, the cross point of $L=14$ is used as the 
crossover point between the NTLL and SDW$_2$TLL regions. 
The product of $\eta_2 \eta_z$ is also plotted in Fig.~\ref{eta}. 
The characteristic condition of the TLL theory $\eta_2 \eta_z=1$ 
is well satisfied around the cross point.

\begin{figure}[ht]
\includegraphics[width=0.85\linewidth,angle=0]{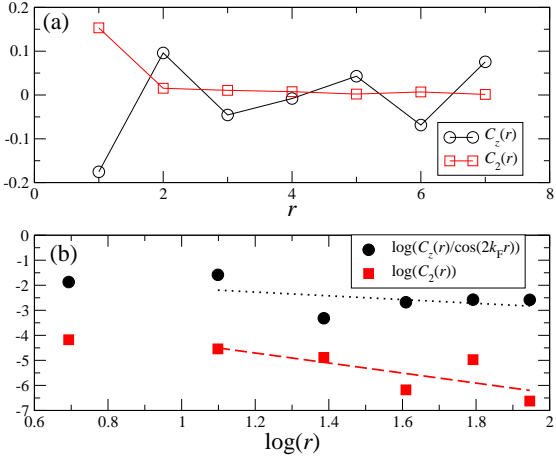}%
\caption{\label{czc2M2} 
Behaviors of $C_z(r)$ and $C_2(r)$ in case of $J_1=0.5$, $\lambda = 2.5$ and $m=2/14$
in the SDW$_2$TLL region.
The dotted and broken lines in (b) are guide for the eye.
We see that $C_z(r)$ is predominant over $C_2(r)$.
}
\end{figure}

\begin{figure}[ht]
\bigskip
\bigskip
\includegraphics[width=0.85\linewidth,angle=0]{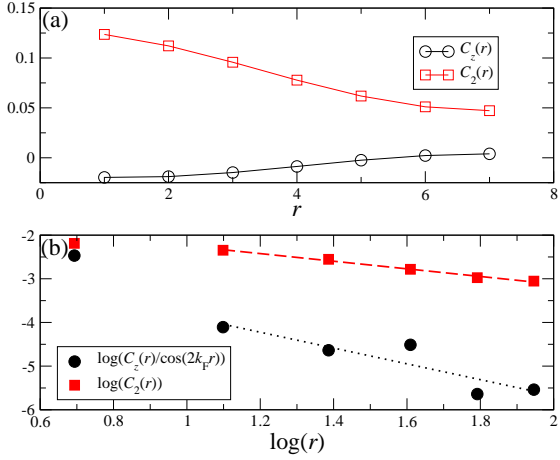}%
\caption{\label{czc2M10} 
Behaviors of $C_z(r)$ and $C_2(r)$ in case of $J_1=0.5$, $\lambda = 2.5$ and $m=10/14$
in the NTLL region.
The dotted and broken lines in (b) are guide for the eye.
We see that $C_2(r)$ is predominant over $C_z(r)$.
}
\end{figure}

We show the behaviors of $C_z(r)$ and $C_2(r)$ in Figs.~\ref{czc2M2} and \ref{czc2M10}.
As already stated,
$C_z(r)$ is predominant over $C_2(r)$ in the SDW$_2$TLL region and vice versa in the
NTLL region.
We can see that this situation is realized in Figs.~\ref{czc2M2} and \ref{czc2M10},
although the precise determination of the critical exponents $\eta_z$ and $\eta_2$
is difficult.
Thus we think that these figures provide the direct confirmation of the  
difference between the SWD$_2$TLL and NTLL regions.
   
\begin{figure}[ht]
\includegraphics[width=0.85\linewidth,angle=0]{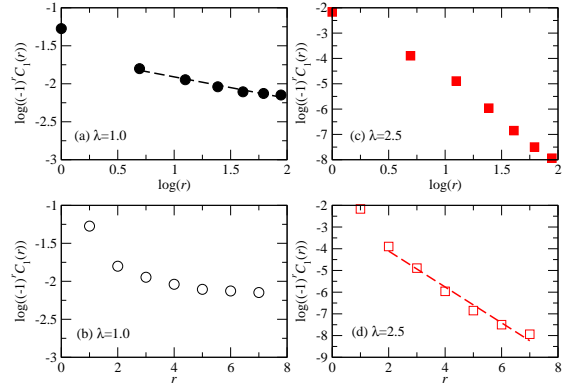}%
\caption{\label{c1M2} 
(a)(b) Behavior of $C_1(r)$ in case of $J_1=0.5$, $\lambda = 1.0$, and $m=2/14$ in the CTLL phase.
(c)(d) Behavior of $C_1(r)$ in case of $J_1=0.5$, $\lambda = 2.5$, and $m=2/14$ in the
SDW$_2$TLL region.
The broken lines are guide for the eye.
}
\end{figure}

\begin{figure}[ht]
\bigskip
\bigskip
\includegraphics[width=0.85\linewidth,angle=0]{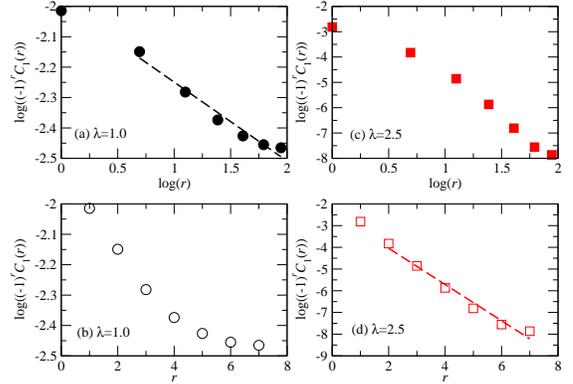}%
\caption{\label{c1M10} 
(a)(b) Behavior of $C_1(r)$ in case of $J_1=0.5$, $\lambda = 1.0$, and $m=10/14$ in the CTLL phase.
(c)(d) Behavior of $C_1(r)$ in case of $J_1=0.5$, $\lambda = 2.5$, and $m=10/14$ in the
NTLL region.
The broken lines are guide for the eye.
}
\end{figure}

We also calculate the two-spin correlation function defined by
\begin{equation}
   C_1(r)
   \equiv \langle S_{\alpha,0}^+ S_{\alpha,r}^- \rangle
\end{equation}
which is expected to decay in the power law in the CTLL phase
while in the exponential law in the two-magnon bound TLL phase.
The behavior of $C_1(r)$ is shown in Figs.~\ref{c1M2} and \ref{c1M10}.
In both figures the magnitudes of $C_1(r)$ are much larger in the CTLL phase
than in the two-magnon bound TLL phase (SDW$_2$TLL and NTLL regions).
Also the decay patterns of $C_1(r)$ are consistent with the above expectation.
Therefore Figs.~\ref{c1M2} and \ref{c1M10} provide the direct confirmation
of the phase transition between the CTLL phase and the two-magnon bound TLL phase.

\section{Phase Diagrams}

\begin{figure}
\includegraphics[width=0.85\linewidth,angle=0]{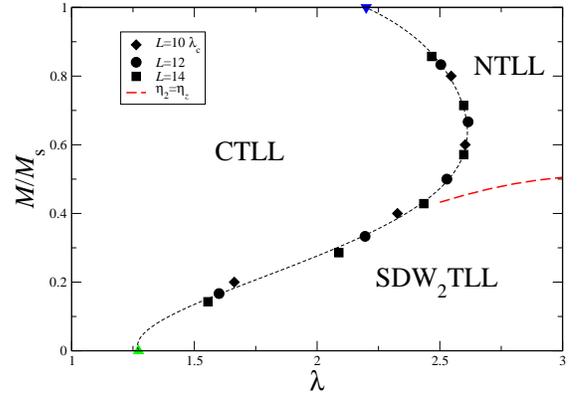}%
\caption{\label{mphasej115} 
Phase diagram with respect to the anisotropy and the magnetization for $J_1=1.5$. 
}
\end{figure}

Now we present several phase diagrams including the field-induced 
NTLL and SDW$_2$TLL regions. 
At first the phase diagrams on the plane of the anisotropy $\lambda$ and 
the reduced magnetization $m=M/M_s$ for $J_1$=0.5, 1.0 and 1.5 are shown in 
Figs.~\ref{mphasej105}, \ref{mphasej110} and \ref{mphasej115}, respectively. 
The phase boundary between the CTLL and the two-magnon-bound 
TLL phases is obtained as the cross of $\Delta _1$ and $\Delta_{2k_{\rm F}}$ 
at each $M$ for $L=$10 (diamond), 12 (circle) and 14 (square). 
The two-magnon-bound TLL phase is divided into the NTLL and SDW$_2$TLL regions 
by the crossover line (broken red curve) determined by $\eta_2=\eta_z$. 
The critical point between the Haldane and N\'eel phases at $M=0$ 
determined by the phenomenological renormalization group method
is given as a green triangle $\blacktriangle$. 
The phase boundary (blue triangle {$\blacktriangledown$})
at $m=M/M_s=1$ is the point where the saturation field changes 
from $H_{\rm s1}=E(L,L)-E(L,L-1)$ to $H_{\rm s2}=[E(L,L)-E(L,L-2)]/2$, which 
is almost independent of $L$. 
The dashed curve is the guide for the eye for the phase boundary between CTLL and 
the two-magnon-bound TLL phases. 

\begin{figure}[h]
\bigskip
\bigskip
\includegraphics[width=0.85\linewidth,angle=0]{ladder-mphasej105cc.eps}%
\caption{\label{mphasej105} 
Phase diagram with respect to the anisotropy and the magnetization for $J_1=0.5$. 
For the green triangle $\blacktriangle$ and blue triangle $\blacktriangledown$,
see the text.}

\bigskip
\bigskip
\includegraphics[width=0.85\linewidth,angle=0]{ladder-mphasej110cc.eps}%
\caption{\label{mphasej110} 
Phase diagram with respect to the anisotropy and the magnetization for $J_1=1.0$. }
\end{figure}

\begin{figure}
\includegraphics[width=0.85\linewidth,angle=0]{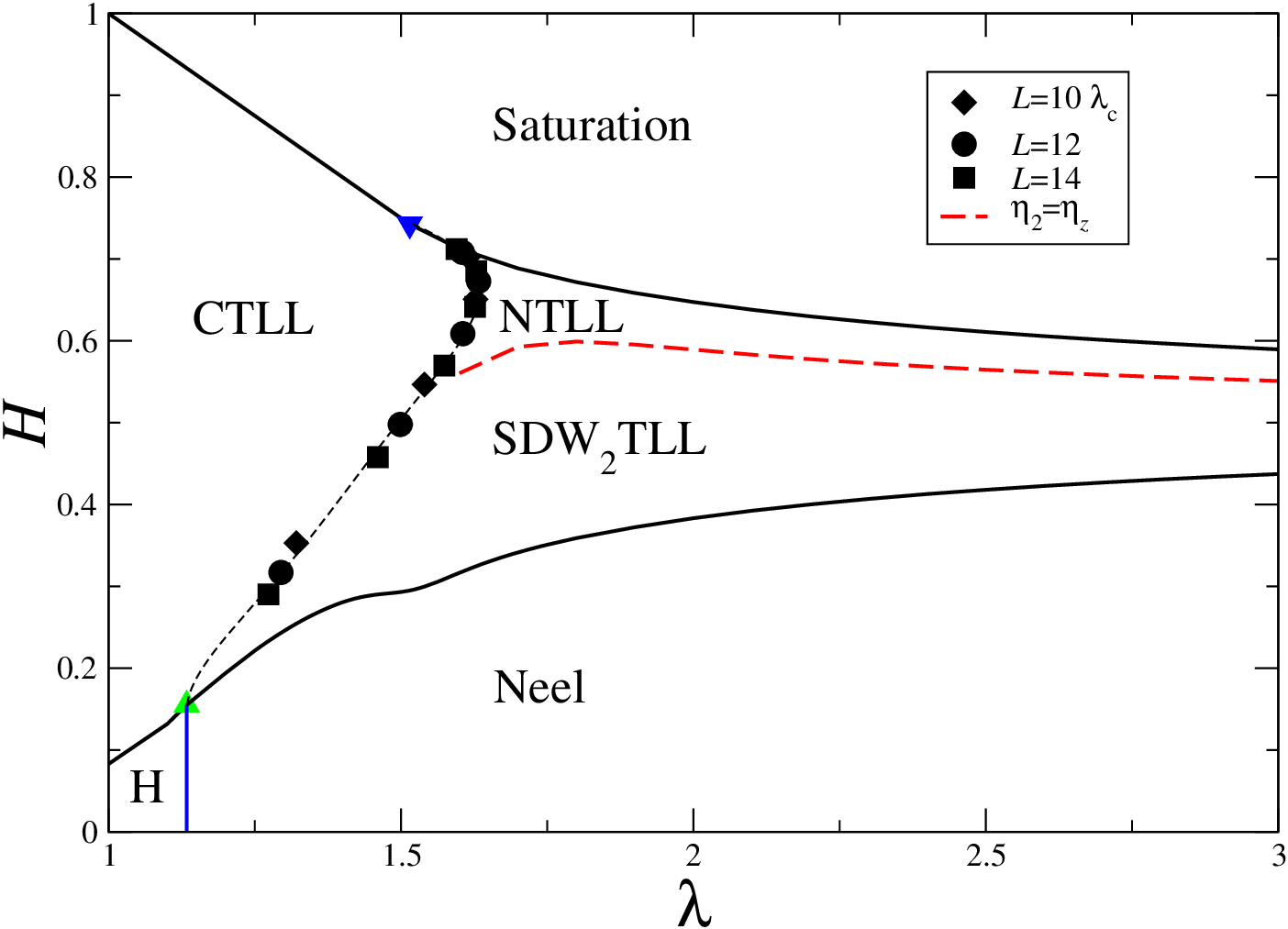}%
\caption{\label{hphasej105} 
Phase diagram with respect to the anisotropy and the magnetic field for $J_1=0.5$. 
The phase denoted by H is the Haldane phase.}
\bigskip
\bigskip
\includegraphics[width=0.85\linewidth,angle=0]{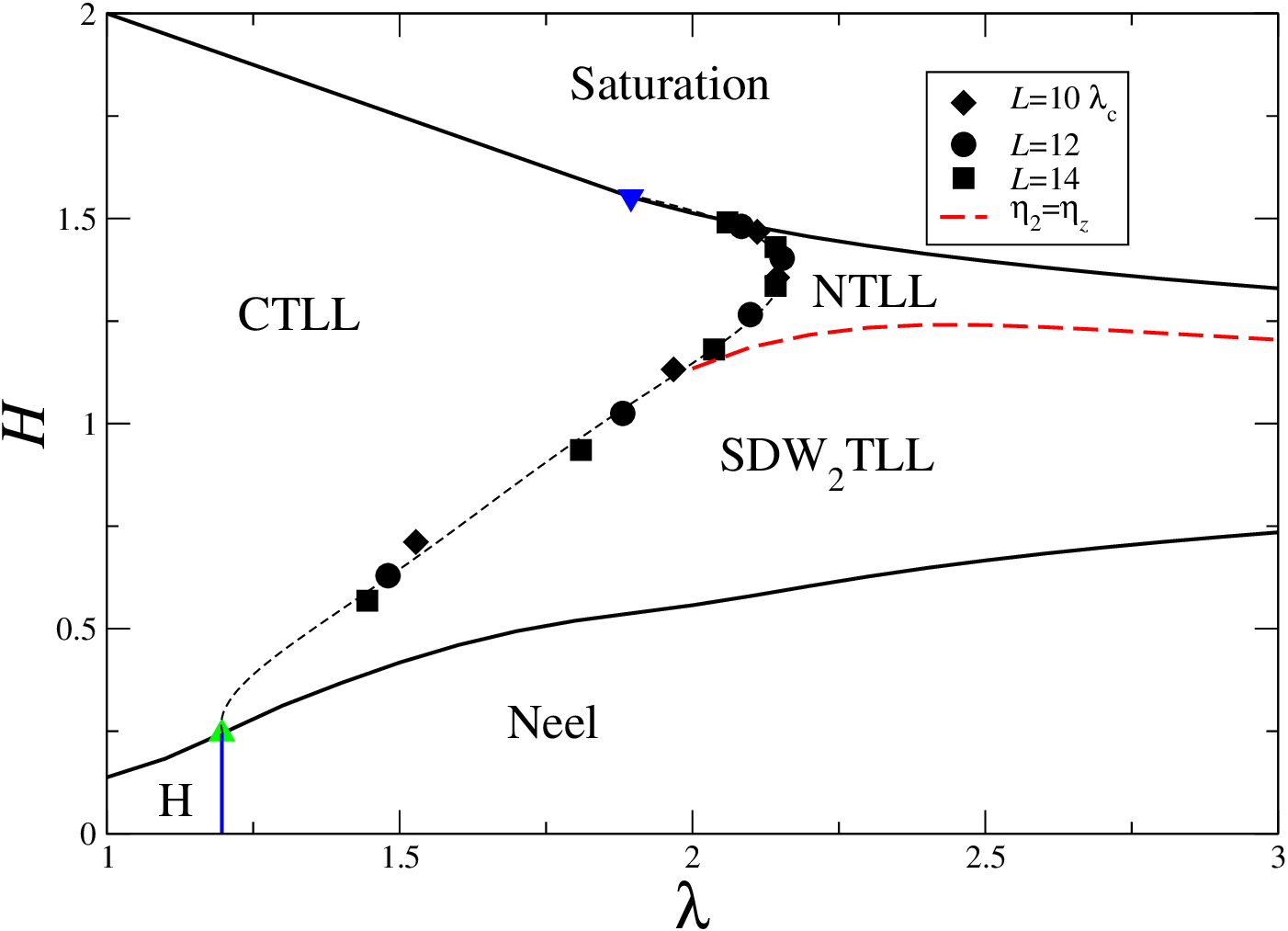}%
\caption{\label{hphasej110} 
Phase diagram with respect to the anisotropy and the magnetic field for $J_1=1.0$. 
}
\bigskip
\bigskip
\includegraphics[width=0.85\linewidth,angle=0]{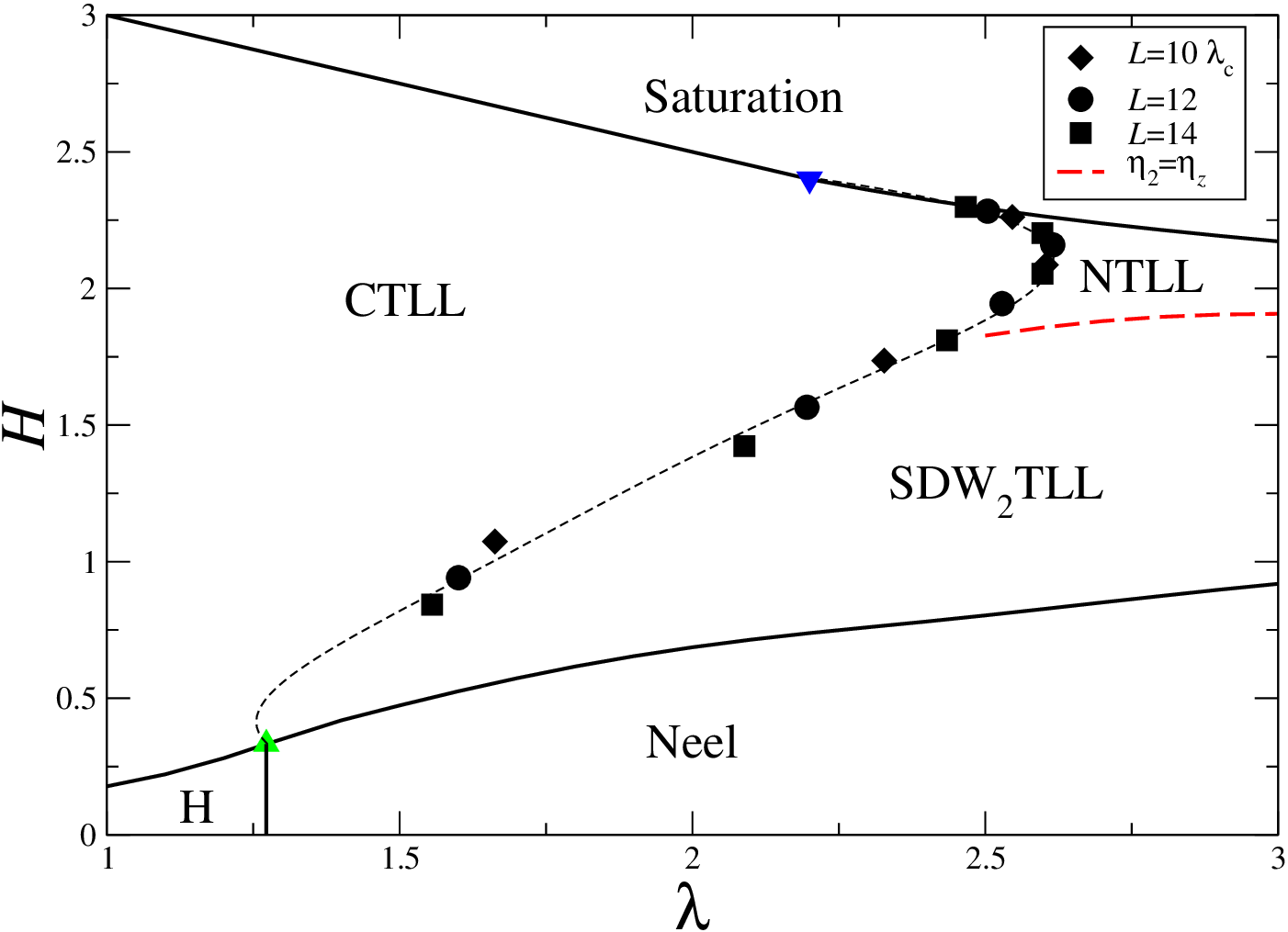}%
\caption{\label{hphasej115} 
Phase diagram with respect to the anisotropy and the magnetic field for $J_1=1.5$. 
}

\end{figure}

The phase diagrams with respect to the external magnetic field $H$ would be 
much more useful for experimentalists. 
The external field $H$ giving the magnetization $M$ is estimated by the form
\begin{eqnarray}
H=[E(L,M+1)-E(L,M-1)]/2
\label{field1}
\end{eqnarray}
at each phase boundary and by the form
\begin{eqnarray}
H=[E(L,M+2)-E(L,M-2)]/4
\label{field2}
\end{eqnarray}
at each crossover point. 
The phase boundary between the Haldane (H) and CTLL phases is estimated by 
Shanks transformation\cite{shanks,barber} applied to the critical field $H_{\rm c1}^{(1)}$ given by 
\begin{eqnarray}
H_{\rm c1}^{(1)}=E(L,1)-E(L,0)
\label{h1}
\end{eqnarray}
calculated for $L=6, 8, 10, 12$ and $14$. 
The boundary between the N\'eel and the two-magnon-bound TLL phases is estimated by 
the same method applied to $H_{\rc 1}^{(2)}$ given by 
\begin{eqnarray}
H_{\rc 1}^{(2)}=[E(L,2)-E(L,0)]/2
\label{h2}
\end{eqnarray}
The Shanks transformation applied to a sequence \{$P_L$\} is 
defined as the form
\begin{eqnarray}
\label{shanks}
P'_{L}={{P_{L-2}P_{L+2}-P_L^2}\over{P_{L-2}+P_{L+2}-2P_L}}.
\end{eqnarray}
The result of this transformation applied to $H_{c2}$ 
for $J_1=0.5$ and $\lambda =2.0$ is shown in Table I.
The phase diagrams in the $\lambda$-$H$ plane for $J_1=0.5, 1.0$ and $1.5$ are shown in 
Figs.~\ref{hphasej105}, \ref{hphasej110} and \ref{hphasej115}, respectively. 
In all the cases the NTLL region appears only in the magnetization process 
from the N\'eel ordered phase.

\begin{table}[h]
   \caption{Result of the Shanks transformation applied to the sequence 
   $H_{c2}=[E(L,2)-E(L,0)]/2$ twice for $J_1=0.5$ and $\lambda =2.0$. }
   \bigskip
   \begin{tabular}{|c|c|c|c|}
      \hline
      $L$& $P_L$ & $P_L'$ &$P_L''$ \\ \hline
      ~6~& ~  0.4320515 & & \\ \hline
      ~8~&    0.4138989 & ~0.3920970~ & \\ \hline
      ~10~&    0.4039936 & 0.3873292 & ~0.3832358~ \\ \hline
     ~12~&   0.3977810 & 0.3851267  & \\ \hline
     ~14~&   0.3936142 && \\ \hline
   \end{tabular}
   \label{shanks1}
\bigskip
\bigskip
\end{table}

\section{Magnetization Curves}

\begin{figure}
\includegraphics[width=0.85\linewidth,angle=0]{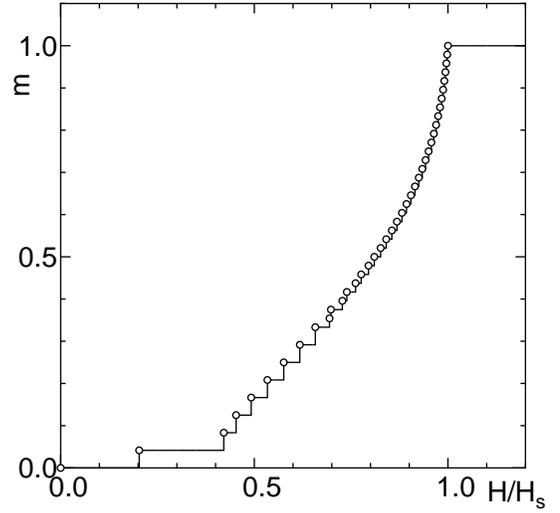}%
\caption{\label{magj105d150} 
Magnetization curve obtained by the DMRG for $J_1=0.5$ and $\lambda =1.50$. 
}
\end{figure}

\begin{figure}
\includegraphics[width=0.85\linewidth,angle=0]{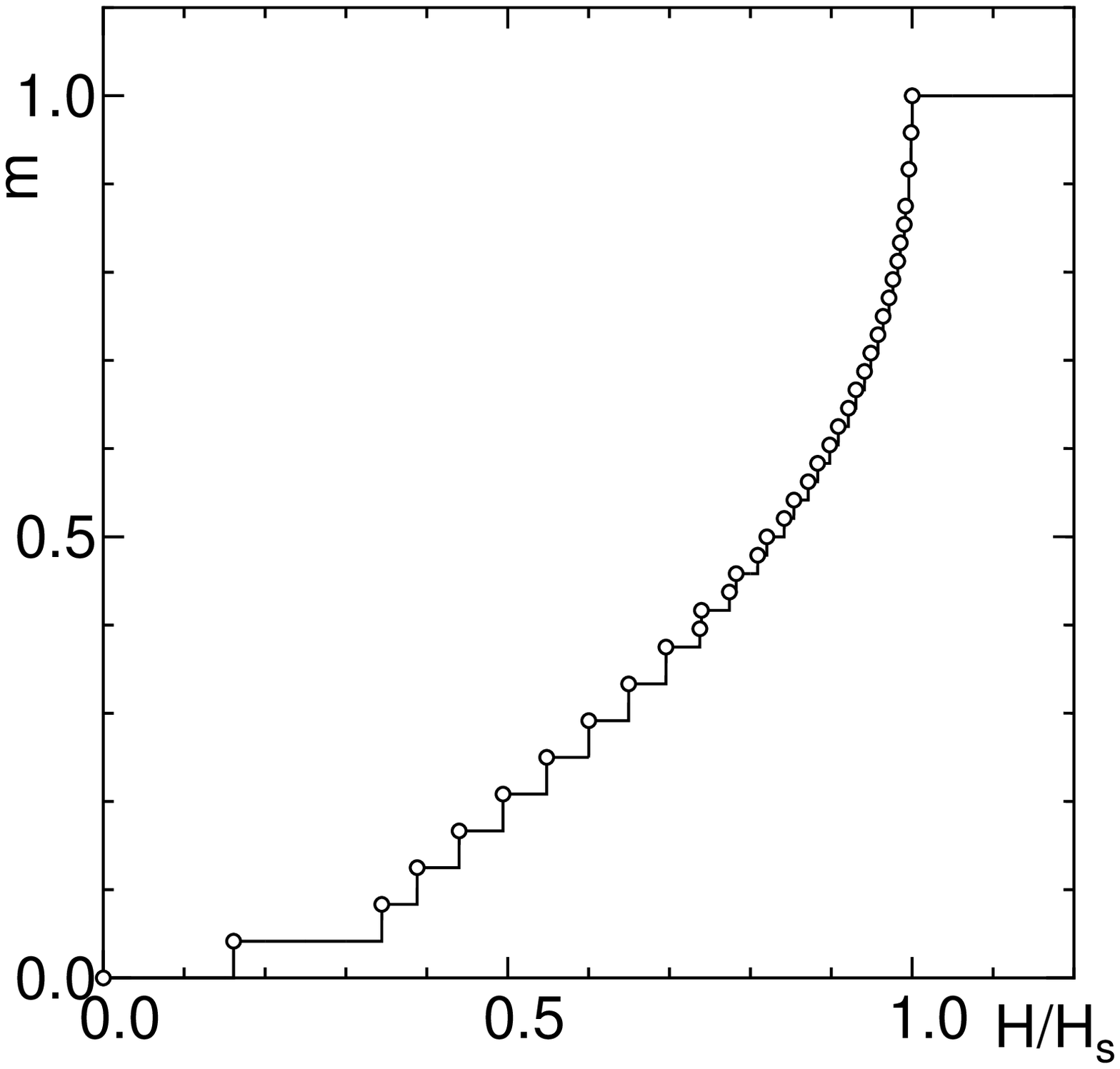}%
\caption{\label{magj105d155} 
Magnetization curve obtained by the DMRG for $J_1=0.5$ and $\lambda =1.55$. 
}
\end{figure}

\begin{figure}
\includegraphics[width=0.85\linewidth,angle=0]{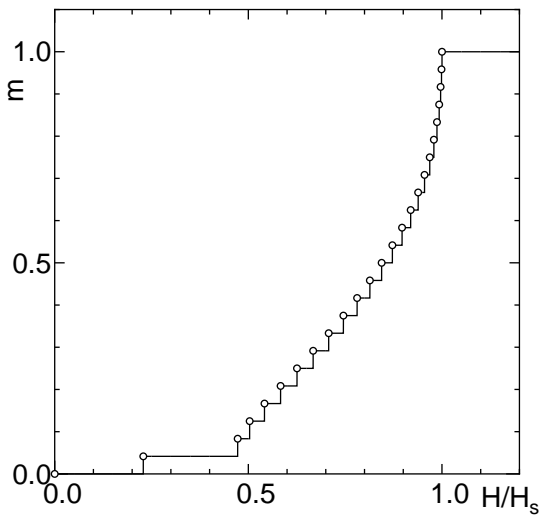}%
\caption{\label{magj105d160} 
Magnetization curve obtained by the DMRG for $J_1=0.5$ and $\lambda =1.60$. 
}
\end{figure}

The magnetization curves are calculated by the DMRG method for $L=96$ 
under the open boundary condition with fixed $J_1=0.5$ for 
$\lambda=1.50, 1.55$ and $1.60$ shown in Figs.~\ref{magj105d150}, 
\ref{magj105d155} and \ref{magj105d160}, respectively. 
The region with $\delta M=1$ corresponds to the CTLL phase, 
while that with $\delta M=2$ the two-magnon-bound TLL phase. 
These three magnetization curves are consistent with 
the phase diagrams in Figs.~\ref{mphasej105} and \ref{hphasej105}. 
In Fig.~\ref{magj105d150} the field-induced transition from 
SDW$_2$TLL to CTLL phases occurs at $H/H_s\sim 0.7$. 
In Fig.~\ref{magj105d155} the first transitions from the SDW$_2$TLL to the 
CTLL phases occurs at $H/H_s\sim 0.7$ and the second one to the 
NTLL phase appears just before the saturation. 
In Fig.~\ref{magj105d160} the crossover from the SDW$_2$TLL to the 
NTLL regions is expected to occur,
which does not bring about any change of the magnetization step. 
At any field induced transition or crossover, 
the magnetization curve does not exhibit any significant anomalous 
behavior, such as the magnetization plateau, jump, or kink etc.

\section{Discussion}

In order to propose the experiments to observe the field induced NTLL phase, 
we should look for the $S=1/2$ spin ladder systems with the ferromagnetic 
rung interaction. 
Possible candidate materials are as follows: 
(CH$_3$)$_2$CHNH$_3$CuCl$_3$,\cite{manaka1,manaka2,masuda,garlea,hong,vaccarelli}
Li$_2$Cu$_2$O(SO$_4$)$_2$\cite{vaccarelli2} and the organic spin ladder 
3-I-V[3-(3-iodophenyl)-1,5-diphenylverdazyl].\cite{yamaguchi} 
Among three compounds the first and the second ones were revealed to 
have the spin gap which suggests the Haldane state at zero magnetic field.
Thus they have no chance to observe the NTLL. 
On the other hand, for the third one the N\'eel order was observed at $H=0$ and 
a kind of multipolar order was observed near the saturation 
of the magnetization process. 
Some more extensive analyses on this material would be interesting. 

In order to detect the spin nematic TLL phase, the NMR measurement 
is one of suitable methods. 
The critical exponent of the spin correlation function $\eta_z$ 
calculated in the section V 
can be estimated from the temperature dependence of the 
NMR relaxation rate.\cite{sato2,sato3} 
The region for $\eta_z > 1$ at low temperature 
is expected to be in the spin nematic TLL phase. 
Actually the spin nematic behavior of the frustrated spin chain compound LiCuSbO$_4$ was 
observed by this experimental method.\cite{grafe} 
The NMR measurement on some spin ladder materials would be also 
quite interesting. 
Furuya\cite{furuya2017} showed that a characteristic angular dependence
of the linewidth of the paramagnetic resonance peak in the ESR absorption spectrum
occurred in the two-magnon-bound TLL phase.
This experiment is also strongly desirable.

\section{Summary}

The $S=1/2$ spin ladder system with the anisotropic ferromagnetic 
rung interaction under magnetic field is investigated using 
the numerical diagonalization for finite-size clusters
and the DMRG analyses. 
The critical exponent analysis reveals that,
in the field-induced two-magnon-bound TLL phase,
the NTLL region appears for large $H$, while the SDW$_2$TLL one for small $H$.  
The magnetization curves calculated by the DMRG indicates that 
after the field-induced phase transition from the SDW$_2$TLL to the CTLL phases, 
the transition to the NTLL phase would possibly occur, for some suitable 
parameters. 
Several phase diagrams with respect to the coupling anisotropy, 
the magnetization and the magnetic field are presented. 
It would be a good proposal of the candidate system that exhibit the 
field-induced spin nematic liquid phase, 
without the frustration or the biquadratic exchange interaction. 

The field-induced nematic liquid phase in unfrustrated system has been found
in $S=1$ models\cite{sakai1998,sakai2022}
and $S=1/2$ ladder model with ferromagnetic rung interactions.\cite{sakai1}
The key point for the realization of the field-induced nematic liquid phase in
the former models is the easy-axis on-site anisotropy which chooses
the $|S^z = \pm 1\rangle$ states over the $|S^z = 0\rangle$ state.
For the latter model the key point is the Ising-like $XXZ$ anisotropy
of the ferromagnetic bond which prefers the $|\uparrow \uparrow \rangle$ and
$|\downarrow \downarrow \rangle$ states
to the $(1/\sqrt{2})|\uparrow \downarrow \pm \downarrow \uparrow \rangle$
states of the spin pair connected by the ferromagnetic interaction.
These key points are essentially the same as each other
as can be seen by mapping the latter model onto the $S=1$ model.
Thus the field-induced nematic liquid phase is expected in models
in which the above key point is satisfied.
In fact, recently we have found the field-induced nematic liquid phase
in the $S=1/2$ ferromagnetic-antiferromagnetic bond-alternating chain\cite{nakanishi}
and in the $S=1/2$ $\Delta$-chain with ferromagnetic interactions.\cite{sakai-sces2022}

\begin{acknowledgments}

This work was partly supported by JSPS KAKENHI, 
Grant Numbers JP16K05419, JP20K03866, JP16H01080 (J-Physics), 
JP18H04330 (J-Physics) and JP20H05274.
A part of the computations was performed using
facilities of the Supercomputer Center,
Institute for Solid State Physics, University of Tokyo,
and the Computer Room, Yukawa Institute for Theoretical Physics,
Kyoto University.
We used the computational resources of the supercomputer 
Fugaku provided by the RIKEN through the HPCI System 
Research projects (Project ID: hp200173, hp210068, hp210127, 
hp210201, and hp220043). 
\end{acknowledgments}


\end{document}